\documentclass[runningheads]{llncs}

\usepackage{graphicx}
\usepackage{multirow}
\usepackage{stfloats}
\usepackage{xcolor}

\usepackage{comment}
\usepackage{todonotes}

\title{Toward Realistic Privacy Risk in FL-Enabled OT Intrusion Detection Systems}

\titlerunning{Toward Realistic Privacy Risk in FL-Enabled OT IDSs}

\author{
Assunta Imperatrice\inst{1,2}
\and
Luigi Romano\inst{2}
}

\institute{
IMT School for Advanced Studies Lucca, Lucca, Italy
\and
Parthenope University of Naples, Naples, Italy
\email{assunta.imperatrice@imtlucca.it, luigi.romano@uniparthenope.it}
}

\authorrunning{A. Imperatrice and L. Romano}
\begin{document}

\maketitle




\enlargethispage{5pt} 
\begin{abstract}

In Operational Technology (OT) environments, Federated Learning (FL) is increasingly adopted to train Intrusion Detection Systems (IDSs) without sharing raw operational data across facilities. However, model updates may still leak sensitive information about industrial processes, creating privacy risks that remain largely invisible within current OT security architectures. Existing reference models, including the Purdue hierarchy, do not explicitly represent the FL learning lifecycle, making privacy leakage difficult to observe and evaluate. To address this gap, we introduce the ML Plane, a novel architectural abstraction that integrates FL workflows within OT systems, enabling visibility of learning-related information flows that are not explicitly represented in traditional OT architectures. Building on the proposed ML Plane, we outline a privacy auditing framework for FL-enabled OT systems. The framework uses Federated Membership Inference Attacks (FedMIA) as an initial assessment mechanism and is being instantiated within a Smart Grid testbed to support the future evaluation of privacy risks under realistic OT deployment constraints.

\vspace{1.5ex}
\end{abstract}

\section{Problem Definition}

Operational Technology (OT) environments increasingly rely on AI-based Intrusion Detection Systems (IDSs) to detect cyber threats and anomalous process behavior. However, the deployment of accurate detection models is constrained by two well-known challenges: the limited availability of labeled attack data and the confidentiality of industrial process information. These constraints often prevent organizations from sharing operational datasets, even when collaborative learning could improve detection capabilities. 
Federated Learning (FL) has emerged as a promising solution to this problem \cite{mcmahan2017fedavg}. By allowing multiple sites to collaboratively train a shared model while keeping raw data within their respective operational domains, FL enables knowledge sharing without centralized data collection \cite{yang2019federated}. For this reason, FL is attracting growing interest across industrial sectors, including manufacturing, energy and critical infrastructures, where data sovereignty and operational confidentiality are the main concerns.
Training generates learning artifact such as model updates and gradients. Previous studies have demonstrated that attacks, such as gradient inversion and Membership Inference Attacks, can successfully extract sensitive information from model updates \cite{zhu2019dlg,shokri2017membership}.
To support collaborative training, updates generated by all participating clients converge at a common aggregation point before the global model is produced. As a result, information originating from multiple independent operational domains becomes observable within a single learning infrastructure. An honest-but-curious aggregator, namely an entity that correctly executes the FL protocol while analyzing the updates it receives, gains access to the unaggregated learning artifacts of all participants before aggregation attenuates their individual characteristics. This creates a unique inference surface spanning the entire federation and enables systematic analysis across multiple clients and operational contexts. This represents a realistic lower-bound adversarial model. Such a setting naturally aligns with Federated Membership Inference Attacks (FedMIA) and the ``all-for-one'' principle introduced in \cite{zhu2025fedmia}, where information leakage can be assessed from the collective observation of federated updates. In OT environments, the inferred information may reveal operational regimes, production cycles, maintenance conditions or process anomalies, exposing knowledge that may facilitate industrial espionage or the preparation of cyber-physical attacks.
Frameworks like Purdue hierarchy and NIST SP 800-82 Rev.3 \cite{stouffer2023} were developed to represent operational communication flow, including sensor measurements, control commands, and supervisory traffic, providing a structured view of industrial communications and trust boundaries. FL introduces a different category of information flows: model updates, gradients, and aggregation exchanges have no explicit representation in these models, even though they routinely traverse the trust boundaries between OT zones and aggregation services located in the DMZ or enterprise environment. Because these exchanges are perceived as ordinary application traffic, the aggregation point and the inference surface it creates fall outside the architectural visibility of existing reference models and cannot be systematically observed or evaluated.
To provide architectural visibility into these learning-specific information flows, we introduce the \textit{ML Plane}, an orthogonal abstraction that explicitly represents the FL lifecycle and its associated artifacts within OT systems. The ML Plane complements traditional OT architectural views by treating learning artifacts as first-class entities, enabling the systematic analysis of information flows generated by distributed learning processes. Through this abstraction, privacy exposure becomes observable as an architectural property of the system rather than as an isolated characteristic of individual machine learning models.
Building upon the ML Plane, we develop a cross-layer privacy auditing framework that evaluates information exposure at the aggregation layer under an honest-but-curious threat model. The framework leverages FedMIA as a representative privacy assessment mechanism and is designed for realistic OT deployments characterized by cyclic and non-IID process data, constrained edge resources, and limited applicability of computationally expensive protections such as Secure Aggregation.
The work presented in this paper describes an ongoing research effort aimed at enabling realistic privacy-risk evaluation in FL-enabled OT intrusion detection systems. In particular, our current research focuses on the following objectives:

\begin{itemize}
    \item \textbf{ML Plane: A cross-layer architectural abstraction.} We introduce the ML Plane, an architectural extension that explicitly represents FL artifacts, including model updates, gradients, and aggregation exchanges, within OT system architectures.
    \item \textbf{Cross-layer privacy auditing framework.} We develop a privacy auditing framework that enables the systematic evaluation of information exposure through FL updates under an honest-but-curious aggregator model.
    \item \textbf{Realistic OT federated learning testbed.} We implement a Smart Grid testbed integrating ICSSIM, NVIDIA FLARE, and industrial communication protocols to support privacy evaluation under realistic OT deployment conditions.
\end{itemize}

\section{ML Plane: a cross-layer architectural abstraction}

\begin{figure}
    \centering
    \includegraphics[width=\columnwidth]{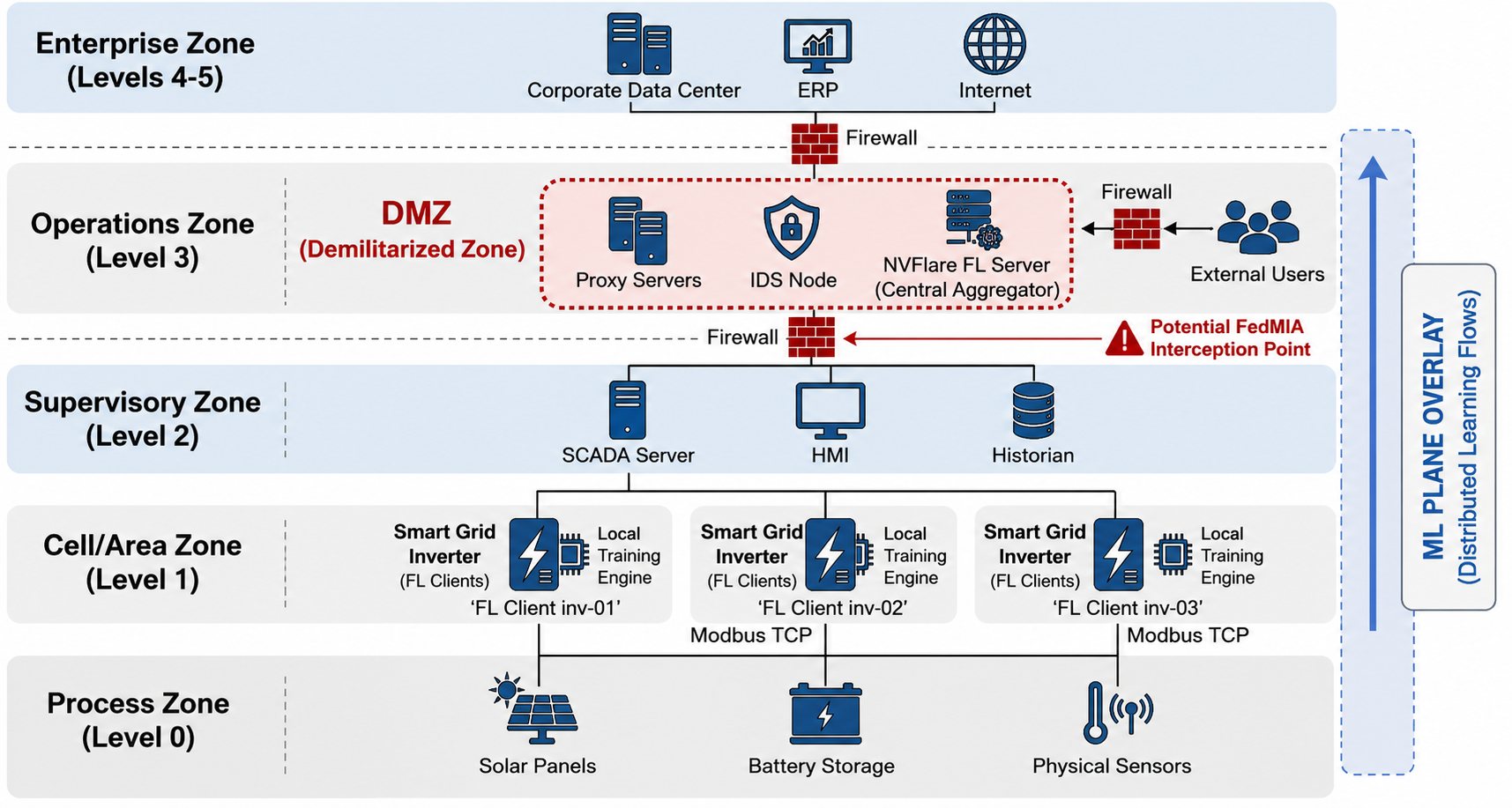}
    \caption{Architectural integration of the ML Plane within the Purdue hierarchy, highlighting the placement of the aggregation-layer observation module.}
    \label{architecture}
\end{figure}

The Purdue model is the de facto reference architecture for network segmentation and security in Operational Technology (OT) environments  \cite{williams1994purdue,stouffer2023}. It organizes industrial infrastructures into a hierarchical multi-layer architecture that separates physical processes, control systems, supervisory functions, and enterprise service through well-defined trust boundaries\cite{iec62443}.
At a high level, Levels 0--2 comprise the industrial process and control layers, Level 3 hosts operational management services, and Level 4--5 correspond to enterprise IT systems. An Industrial Demilitarized Zone (DMZ) is typically deployed between operational and enterprise domains to regulate inter-zone communications and reduce attack propagation \cite{stouffer2023}.
However, the introduction of FL creates information flows that fall outside the visibility provided by the Purdue model. 
During federated training, edge nodes exchange high-dimensional gradient vectors and weight matrices with a central aggregator located in the DMZ, as shown in Figure \ref{architecture}. In contrast, the FL workflow follows a client-to-aggregator axis that spans multiple Purdue levels and crosses several trust boundaries without belonging to any specific architectural zone. As a result, perimeter defenses can monitor the communication channel, but they provide no visibility into the privacy-sensitive information embedded within the model parameters.

To provide architectural visibility into these learning-related information flows, we introduce the ML Plane, an orthogonal architectural abstraction that spans the Purdue hierarchy and explicitly represents the lifecycle of FL artifacts. Rather than introducing a new Purdue level, the ML Plane provides an orthogonal view of learning-related information flows that coexist with traditional operational communications.
Within the ML Plane, physical assets such as solar panels, battery storage systems, and sensors generate operational measurements in the Process Zone (Level 0). The Cell/area Zone (Level 1) devices, such as smart inverters, act as FL clients that locally train models and produce parameter updates. Each client acquires its local process data via Modbus TCP and runs a Local Training Engine that transforms it into a model update. These updates do not remain at the edge, but traverse supervisory layers and cross the DMZ toward the aggregation server, where they are combined into a global model and redistributed to participating clients.  Zone-based controls, such as firewalls, intrusion prevention systems, and the DMZ can observe the presence of model-update exchanges but remain structurally blind to the information content encoded within them. In a single training round, the learning flow traverses multiple Purdue levels and trust boundaries between the edge nodes and the DMZ. As illustrated in Figure \ref{architecture}, the aggregation server therefore becomes a natural intercept point for FedMIA-based privacy analysis. 

\section{Cross-layer privacy auditing framework}

\begin{figure}[t]
    \centering
    \includegraphics[width=\columnwidth]{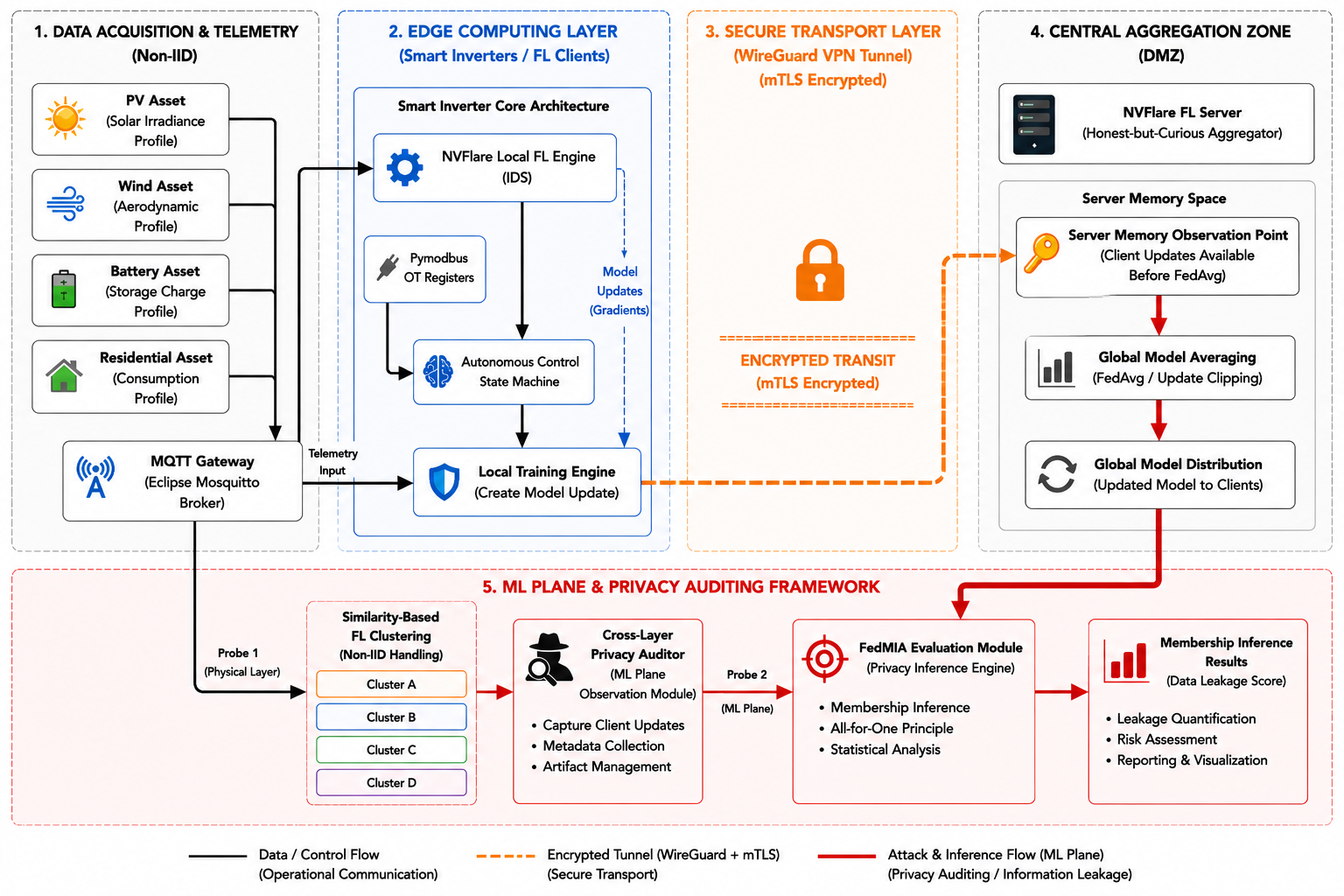}
    \caption{End-to-end experimental workflow instantiated within the ML Plane.}
    \label{fig: Pipeline}
\end{figure}

To evaluate privacy risks in FL-enabled OT environments, we develop a cross-layer privacy auditing framework that operationalizes the ML Plane introduced in the previous section. The framework is designed to systematically assess the information exposed through FL workflows and is structured around three key components: a threat model, an attack surface, and an observation mechanism. Figure~\ref{fig: Pipeline} illustrates the end-to-end auditing workflow. 

\textbf{Threat Model, Attack Surface and an Observation Mechanism}. We consider an honest-but-curious aggregator that executes the FL protocol while having access to all model updates exchanged during training, with the goal of inferring sensitive information from them. This represents a realistic lower-bound adversarial model for privacy risk evaluation in federated learning systems deployed in OT environments.
The attack surface is defined by the federated learning artifacts exposed through the ML Plane (Fig.~\ref{architecture}), namely local model updates and aggregated global models. Although these artifacts are not visible within traditional OT reference architectures, they become observable at the aggregation layer, where client updates are temporarily available prior to FedAvg aggregation. This enables an inference surface that can be exploited by membership inference attacks.
To operationalize this setting, we introduce a Privacy Auditor module co-located with the FL aggregation server. In particular, in non-secure aggregation settings, client updates are temporarily available in server memory before the execution of FedAvg. The auditor captures these updates and evaluates privacy leakage using Federated Membership Inference Attacks (FedMIA), comparing inference outcomes against ground-truth process data. This enables a systematic assessment of how learning artifacts encode sensitive operational information.
Figure \ref{fig: Pipeline} illustrates the end-to-end privacy auditing workflow instantiated within the ML Plane. The workflow is organized into five stages. (1) The Data Acquisition and Telemetry stage collects heterogeneous OT measurements from distributed energy assets, generating non-IID operational data streams. (2) The Edge Computing Layer hosts the FL clients, where local IDS models are trained and transformed into model updates. (3) The Secure Transport Layer protects the exchange of learning artifacts through WireGuard and mTLS-protected communication channels. (4) The Central Aggregation Zone performs federated aggregation through the FL server, while exposing a temporary observation point where client updates are available before FedAvg execution. Finally, (5) the ML Plane and Privacy Auditing Framework provides a cross-layer view of the learning workflow, collecting learning artifacts and metadata, applying FedMIA-based analysis, and quantifying potential information leakage. Through this architecture, privacy exposure can be evaluated at the aggregation layer independently of network-level protections.

\textbf{Relation to Existing FL Privacy Research}. Existing research on FL privacy focuses primarily on attack effectiveness and defense mechanisms, including gradient inversion \cite{zhu2019dlg}, membership inference \cite{shokri2017membership,zhu2025fedmia}, differential privacy, and secure aggregation protocol \cite{bonawitz2017practical}. In contrast, our objective is not to propose a new attack or mitigation technique, but to introduce an architectural abstraction that enables privacy auditing within OT reference models. The ML Plane complements existing FL privacy research by providing visibility into learning artifacts that are not explicitly represented in traditional OT architectures.
Although prior studies typically analyze privacy risks at the machine learning layer, treating FL as an isolated learning process, our work focuses on how such risks can be represented, observed, and evaluated within OT system architectures, where learning workflows coexist with industrial control and communication infrastructures. By explicitly mapping learning artifacts onto OT trust boundaries and architectural components, the ML Plane enables privacy leakage to be analyzed as a system-level property emerging from the interaction between FL workflows and industrial operational environments.

\textbf{Realistic OT federated learning testbed}.To evaluate the proposed privacy auditing framework under realistic deployment conditions, we are implementing a Smart Grid FL testbed that reproduces several constraints commonly encountered in OT environments. The testbed models a distributed energy ecosystem composed of heterogeneous assets operating under limited computational resources, intermittent connectivity, and highly non-IID data distributions. 
The infrastructure is built using Containerlab, which enables network emulation and the interconnection of heterogeneous OT components within a controlled environment. Each component is deployed as an independent Docker container. Federated training is orchestrated through NVIDIA FLARE \cite{roth2022nvidia}, while industrial communications are reproduced through standard OT protocols, including Modbus TCP and OPC-UA.  The current implementation adopts a supervised learning workflow in which each client trains a local regression model using its own operational measurements and the corresponding target variables. 
Communications between clients and the aggregation server are protected through WireGuard VPN tunnels and mutual TLS (mTLS), ensuring that privacy leakage is evaluated at the federated learning layer rather than through network-level vulnerabilities.
The energy production and consumption profiles are derived from the Pecan Street dataset, while the operational measurements of the wind turbine are obtained from the Kelmarsh Wind Farm SCADA dataset.
To model non-IID conditions, clients are grouped based on similarity of local distributions, enabling multiple FL clusters representing distinct operational contexts.
This design choice allows direct observation of learning artifacts at the aggregation layer and establishes a baseline for evaluating privacy leakage before introducing additional protection mechanisms. 

\section{Conclusion}

This work introduced the ML Plane, an architectural abstraction designed to expose federated learning workflows within OT environments and to support the systematic analysis of privacy risks associated with distributed model training. Building on this abstraction, we outlined a privacy auditing framework based on an honest-but-curious aggregator model and described its ongoing implementation within a Smart Grid federated learning testbed.
The current effort focuses on completing the integration of the Privacy Auditor and conducting controlled FedMIA experiments to quantify membership-related information leakage under realistic OT deployment constraints. Future work will investigate additional privacy attacks, stronger adversarial models, and the impact of privacy-preserving mechanisms such as Secure Aggregation and Differential Privacy.

\enlargethispage{2\baselineskip}
\bibliographystyle{splncs04}
\bibliography{biblio}
\end{document}